# INFLUENCE OF AN ANISOTROPIC CRYSTALLINE FIELD ON THE MAGNETIC PROPERTIES OF A MIXED SPIN–1/2 AND SPIN–1 ISING MODEL

Jozef STREČKA, Michal JAŠČUR
Department of Theoretical Physics and Astrophysics, Institute of Physics, P. J. Šafárik University, Moyzesova 16,
041 54 Košice, Slovak Republic, tel. 095/6222121–23, E-mail: jozkos@pobox.sk, jascur@kosice.upjs.sk

**SUMMARY**
*The mixed spin–1/2 and spin–1 Ising model in the presence of an anisotropic crystalline field is treated exactly within the framework of an extended star–triangle mapping transformation. The exact results for the phase diagrams, magnetization, internal energy and specific heat are derived and discussed in detail. The relevant mapping suggests that an isotropic in–plane crystal field ($D^x = D^y$) leads to the same effects as the hard–axis crystal field ($D^z$), whereas the in–plane anisotropy ($D^x \neq D^y$) is responsible also for the randomization of the magnetic ordering (transverse–field like effect).*

**Keywords:** exact results, single–ion anisotropy, Ising model, transverse field.

## 1. INDRODUCTION

For many years, the two–sublattice mixed spin–1/2 and spin–1 Ising models have attracted considerable attention, since they are well adopted for the investigation of a certain kind of ferrimagnetism. In particular, the magnetic properties of the mixed spin–1/2 and spin–1 Ising models with a crystal field interaction have been explored by a variety of techniques [1], on the honeycomb lattice even by some exact methods [2]. Among other matters, the strong interest in these models arises partly on account of an interesting critical behaviour they display and partly on account of the possible existence of the compensation phenomenon. However, as far as we know, the most of the theoretical works have been restricted to the investigation of the models with an uniaxial crystal field interaction, whereas the role of a spatially anisotropic crystalline field has not been extensively examined yet. Nevertheless, a number of experimental works revealed that the real magnetic materials often possess a strong source of the anisotropy, such as the a crystalline field, which could be because of the lattice distortion spatially anisotropic [3]. Therefore, the main purpose of this work is to clarify the role of the spatially anisotropic crystalline field on the magnetic properties of the two–sublattice Ising model. It should be mentioned that the relevant physical effects will be treated exactly on the basis of an extended star–triangle transformation. This method is based on the mapping of the mixed honeycomb lattice to its equivalent simple spin–1/2 Ising model on the triangular lattice.

The outline of the present paper is as follows. The fundamental framework of the transformation method is presented in Section 2. The Section 3 deals with the most interesting numerical results and physical consequences of the mapping transformation. Finally, some concluding remarks are given in Section 4.

## 2. FORMULATION

In this work, we will study the mixed–spin 1/2 and 1 Ising model on the honeycomb lattice, in a presence of the spatially anisotropic crystalline field. Let us assume that the sites of the honeycomb lattice occupied by the atoms with spin 1/2 constitute the sublattice A, while the sites that are occupied by atoms with spin 1 constitute the sublattice B. Taking into account the effect of the spatially anisotropic crystalline field on the atoms of sublattice B, the total Hamiltonian of the system reads

$$H_h = -J \sum_{i,j} S_i^z \mu_j^z - D^x \sum_i (S_i^x)^2 - D^y \sum_i (S_i^y)^2 - D^z \sum_i (S_i^z)^2, \quad (1)$$

where the first summation is carried out over the nearest–neighbours only, the summation in other terms are taken over the all sites of sublattice B, $S_i^\alpha (\alpha = x, y, z)$ and $\mu_j^z$ denote the components of the spin–1 and spin–1/2 operators, respectively. Finally, $J$ represents the nearest–neighbour exchange term (considering only the ferrimagnetic case $J<0$) and $D^\alpha (\alpha = x, y, z)$ is the spatially anisotropic single–ion interaction effecting the atoms of sublattice B. For convenience, we can rewrite the total Hamiltonian (1) as a sum of $N/2$ commuting site Hamiltonians $H_k$ ($N$ - a total number of atoms)

$$H_h = \sum_{k=1}^{N/2} H_k, \quad (2)$$

where the each site Hamiltonian $H_k$ is associated with all the interaction terms involving the $k$th atom of sublattice B

$$H_k = -S_k^z E_k - D^x (S_k^x)^2 - D^y (S_k^y)^2 - D^z (S_k^z)^2,$$
$$E_k = J\left(\mu_{k1}^z + \mu_{k2}^z + \mu_{k3}^z\right). \quad (3)$$

Regarding the definition of the site Hamiltonian $H_k$ and the validity of the standard commutation relations $[H_k, H_j] = 0$ (if $k \neq j$), it is possible to write the partition function $Z_h$ of the honeycomb lattice in the following form

$$Z_h = \mathrm{Tr}\exp(-\beta H) = \mathrm{Tr}_{\{\mu\}}\prod_{k=1}^{N/2}\mathrm{Tr}_{S_k}\exp(-\beta H_k), \quad (4)$$

where $\beta = 1/k_B T$, a symbol $\mathrm{Tr}_{\{\mu\}}$ means the trace over the degrees of freedom of atoms of sublattice A and the symbol $\mathrm{Tr}_{S_k}$ represents the trace over the spin states of $k$th atom of sublattice B. In view of further manipulation, it is useful to define new variables $\Delta_1$ and $\Delta_2$ as follows

$$\Delta_1 = (D^x + D^y)/2 \quad \text{and} \quad \Delta_2 = (D^x - D^y)/2. \quad (5)$$

After a straightforward diagonalization of the site Hamiltonian $H_k$, the expression for the partial trace over the spin states of $k$th atom of sublattice B implies the possibility to introduce an extended star–triangle transformation [4]. Actually, the relevant mapping takes, in terms of the new notation (5), the following form

$$\mathrm{Tr}_{S_k}\exp(-\beta H_k) = \exp(2\beta\Delta_1) +$$
$$+ 2\exp[\beta(D^z + \Delta_1)]\cosh\left(\beta\sqrt{E_k^2 + \Delta_2^2}\right) = \quad (6)$$
$$= A\exp\left[\beta R(\mu_{k1}^z \mu_{k2}^z + \mu_{k2}^z \mu_{k3}^z + \mu_{k3}^z \mu_{k1}^z)\right],$$

where the parameters $A$ and $R$ are the unknown star–triangle transformation parameters. Following the standard procedure [4], one can directly obtain the transformation parameters $A$ and $R$

$$A = \exp(2\beta\Delta_1)(V_1 V_2^3)^{1/4}, \quad \beta R = \ln(V_1/V_2). \quad (7)$$

In above, we have defined the functions $V_1$ and $V_2$ in order to express the transformation parameters $A$ and $R$ in more abbreviated and elegant form

$$V_1 = 1 + 2\exp[\beta(D^z - \Delta_1)]\cosh\left(\beta\sqrt{(3J)^2 + (2\Delta_2)^2}/2\right),$$
$$V_2 = 1 + 2\exp[\beta(D^z - \Delta_1)]\cosh\left(\beta\sqrt{J^2 + (2\Delta_2)^2}/2\right). \quad (8)$$

Now, after substituting the extended star–triangle transformation (6) into the expression (4), one can simply derive the relationship between the partition function $Z_h$ of the model under consideration and the partition function $Z_t$ of the triangular lattice with the exchange integral $R$

$$Z_h = A^{N/2} Z_t(\beta, R). \quad (9)$$

Obviously, the above equality represent an essential result of our calculation, since from here onward, all the other thermodynamic quantities can be obtained in the straightforward manner. However, the derivation of some other physical quantities becomes in practice very complex and tedious. Fortunately, we can avoid this problem by exploiting the following exact relations that may be derived after an elementary algebra from the equality (9)

$$\langle f_1(\mu_k^z, \mu_j^z, ..., \mu_l^z)\rangle_h = \langle f_1(\mu_k^z, \mu_j^z, ..., \mu_l^z)\rangle_t,$$
$$\langle f_2(S_k^z, \mu_{k1}^z, \mu_{k2}^z, \mu_{k3}^z)\rangle_h =$$
$$= \left\langle \frac{\mathrm{Tr}_{S_k} f_2(S_k^z, \mu_{k1}^z, \mu_{k2}^z, \mu_{k3}^z)\exp(-\beta H_k)}{\mathrm{Tr}_{S_k}\exp(-\beta H_k)}\right\rangle_t. \quad (10)$$

In equation (10), $f_1$ represents a function depending exclusively on the spins of sublattice A, whereas the function $f_2$ is an arbitrary function depending on the $k$th spin of sublattice B and its three nearest–neighbouring spins of sublattice A. The symbols $\langle ...\rangle_h$ and $\langle ...\rangle_t$ mean the standard ensemble averages related to the honeycomb and triangular lattice, respectively. Applying one of the standard method (e. g. differential operator technique [5]), one can easily obtain the sublattice magnetizations

$$m_A = \langle \mu_k^z\rangle_h = \langle \mu_k^z\rangle_t,$$
$$m_B = \langle S_k^z\rangle_h = \frac{3}{2}(K_1 + K_2)\langle \mu_k^z\rangle_t + \quad (11)$$
$$+ 2(K_1 - 3K_2)\langle \mu_{k1}^z \mu_{k2}^z \mu_{k3}^z\rangle,$$

where the coefficients $K_1$ and $K_2$ are given by

$$K_1 = \frac{6J\exp(\beta\Delta_1 - \beta D^z)}{V_1 \sqrt{9J^2 + 4\Delta_2^2}}\sinh\left(\beta\sqrt{9J^2 + 4\Delta_2^2}/2\right),$$
$$K_2 = \frac{2J\exp(\beta\Delta_1 - \beta D^z)}{V_2 \sqrt{J^2 + 4\Delta_2^2}}\sinh\left(\beta\sqrt{J^2 + 4\Delta_2^2}/2\right). \quad (12)$$

In order to complete the calculation of the spontaneous magnetization of the mixed–spin system on the honeycomb lattice, one must also utilise the well–known results for the spontaneous magnetization and the triplet correlation function of the triangular lattice. Finally, we should mention that the standard thermodynamic approach enables also simple calculation of the internal energy and the specific heat. For the sake of simplicity, we only quote that the calculation of the internal energy $U_h$ and the specific heat $C_h$ of the mixed–spin system on the honeycomb lattice have been made by the use of the standard thermodynamic relations

$$U_h = -\frac{1}{Z_h}\frac{\partial Z_h}{\partial \beta} \quad \text{and} \quad C_h = \frac{\partial U_h}{\partial T}. \qquad (13)$$

## 3. NUMERICAL RESULTS AND DISCUSSION

Before discussing the most interesting numerical results, let us take a closer look at the transformation formulas as given by equation (7) and (8). As it is clearly seen, in the fully isotropic case of the crystal field $D^x = D^y = D^z$ ($\Delta_1 = D^z$, $\Delta_2 = 0$), the effect of the single–ion anisotropy vanishes, since there is no source of the anisotropy. It is also worth mentioning that the isotropic in–plane crystal field (i. e., $D^x = D^y \neq D^z$; $\Delta_1 = D^x, \Delta_2 = 0$) enters in the transformation formulas the same terms, however, as the uniaxial anisotropy $D^z$ with the opposite sign. Consequently, one may conclude that the isotropic in–plane crystal field ($D^x = D^y$) is nothing but the negative uniaxial anisotropy. Finally, another interesting case appears if we consider an extremely anisotropic in-plane crystal field in the system ($D^x \neq 0, D^y = 0, D^z \neq 0$; $\Delta_1 = \Delta_2 = D^x/2$). In this case, the comparison of our results with those of the transverse Ising model on the same system [6] indicates, that the half of the anisotropy $D^x$ simulates a transverse–field like effect (the term $\Delta_2$) and another half acts as a negative uniaxial anisotropy (the term $\Delta_1$). Obviously, any other case ($D^x \neq D^y \neq D^z$) will exhibit both contributions, i. e. the transverse–field like effect (determined by the term $\Delta_2$), as well as the uniaxial single–ion anisotropy effect (determined by the difference $D^z - \Delta_1$).

Now, the numerical results of several particular cases will be discussed. At first, the influence of the isotropic in–plane crystalline field ($D^x = D^y$) will be explored in detail. In Fig. 1 we display the phase boundaries as a function of the isotropic in–plane crystal field for some typical values of the uniaxial anisotropies $D^z$. One observes here that the critical temperature monotonically decreases with the crystal field increasing and it vanishes at the boundary value $D_B^x/|J| = 1.5 + D^z/|J|$, above which only disordered phase may exist. These results are completely consistent with those for the transition temperature dependences on the uniaxial anisotropy $D^z$, since they are only reversed to the positive values of the in–plane crystalline fields $D^x = D^y$. Moreover, it turns out that such a system cannot exhibit the compensation phenomenon. Indeed, in order to demonstrate the overall dependences of the total magnetization $m = (m_A + m_B)/2$, we have depicted in Fig. 2 the thermal variations of the total magnetization for different in–plane crystal fields. The differences in the thermal dependences of the total magnetization arising as a result of the fact, that the magnetization of sublattice B decreases for sufficiently strong in–plane crystal fields more rapidly than the magnetization of sublattice A. Nevertheless, the ground state value of the total magnetization takes its maximum value, what means that the system remains perfectly ordered in the ground state if the anisotropy $D^x$ is less than the boundary value $D_B^x$. Finally, the specific heat variations with the temperature are shown for the same crystal fields in Fig. 3. In addition to the standard Onsager–type behaviour (the curves labelled $D^x/|J|=-2.0$ and 0.0 in the inset of the Fig. 3), an unexpected behaviour in the specific heat can be also found here (see the curves labelled $D^x/|J|=1.0$ and 1.25).

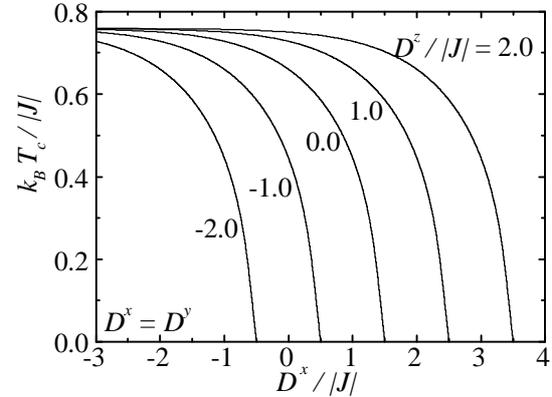

**Fig.1** The transition temperature against anisotropy $D^x/|J|$ for various anisotropies $D^z/|J|$.

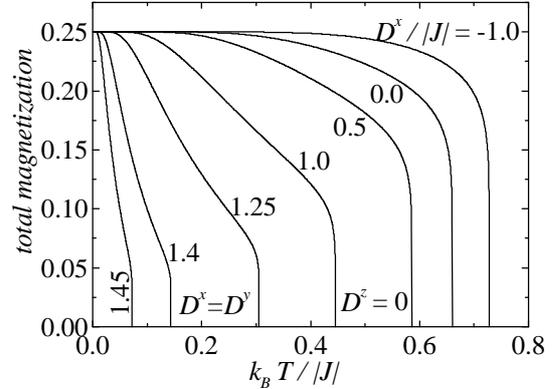

**Fig.2** Thermal variations of the total magnetization when the value of the isotropic in–plane crystalline field is changed and the uniaxial anisotropy $D^z = 0$.

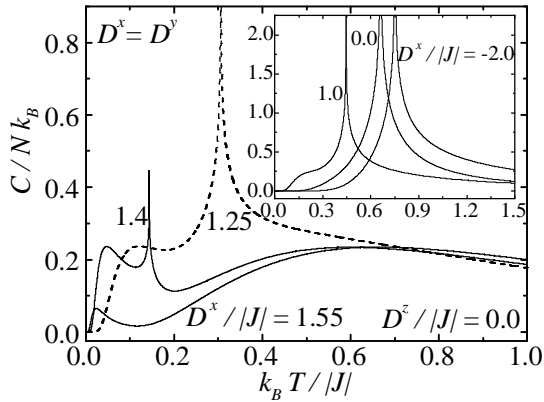

**Fig.3** The specific heat versus temperature for the same crystal fields as in Fig.2. In the inset we have depicted also standard Ising dependences.

In these cases one really finds an anomalous maximum in the low–temperature region of the specific heat. This effect arises due to the crystal field induced thermal reshuffling. Moreover, if the crystal field is sufficiently close to the boundary value $D^x_B$ (see the curve labelled $D^x/|J|=1.4$), the dependence exhibits apart from the low–temperature anomaly another Shottky–type anomaly, in the high–temperature tail of the specific heat. The both maxima clearly penetrating also into the paramagnetic region (the curve $D^x/|J|=1.55$).

Now, we turn our discussion to the solution for the extremely strong in-plane anisotropy ($D^x \neq 0, D^y = 0, D^z \neq 0$), in order to confirm the transverse–field like effect of the crystalline field. Firstly, we illustrate in Fig.4 the results for the phase boundaries (solid lines) and the compensation temperatures (dashed lines). As one can see, the critical frontier strongly depends on whether the anisotropy $D^z$ is positive or negative. Namely, in the former case the ground state remains ordered regardless of the crystal field strength $D^x$ and $D^z$, while in the latter one the transition lines terminate at certain values of the crystal field $D^x$ above which only a disordered phase may occur. It is also easy to understand that the transition lines merge for strong enough negative anisotropies $D^x$, since in this region our system is equivalent to that of large positive uniaxial anisotropy $D^z$ inserted into the strong transverse field. Furthermore, it is noteworthy that the maximum value of the transition temperature is shifted towards more negative anisotropies $D^x$ with the anisotropy $D^z$ decreasing. This effect appears because the relevant negative single–ion anisotropy $D^x$ term preferably compensates the influence of the uniaxial anisotropy $D^z$ and only then the transverse–field like contribution of the crystal field $D^x$ (which causes the repeatedly decrease of the transition temperature) is shown. Another interesting fact to observe here is the compensation phenomenon that occurs for the positive as well negative crystal fields $D^x$. The compensation curve in the region of the negative crystal fields $D^x$ is almost insensitive to $D^x$ and it behaves completely independently of the anisotropy strength $D^z$. Contrary to this, the compensation behaviour in the region of the positive anisotropies $D^x$ is much more complicated, namely, the compensation temperature may increase as well as decrease with increasing in $D^x$, according to anisotropy strength $D^z$.

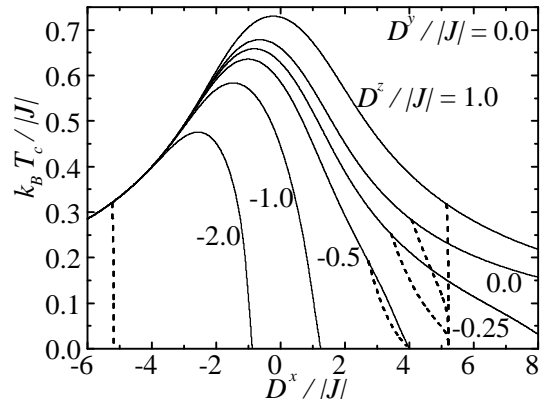

**Fig.4** Phase diagram in the $k_B T_c/|J|-D^x/|J|$ plane, when the anisotropy $D^z$ is changed.

To illustrate the overall thermal dependences of the magnetization, we have depicted in Fig. 5 the total magnetization against the temperature for different crystalline fields $D^x$, when $D^y = 0$ and $D^z = 0$. As it can be seen, the ground state value of the total magnetization does not take always its maximum value (0.25), although it is independent of the sign of the anisotropy $D^x$. Hence, one may conclude that the spontaneous ordering does not depend on the sign of the crystal field interaction $D^x$, even if the transition temperatures differs very much. Moreover, a more detailed study of the sublattice magnetization indicates that the crystal field interaction $D^x$ competes with the exchange interaction, the competition resulting in the randomization of the spontaneous ordering of sublattice B (sublattice A is not directly affected by the crystal field interaction). As a consequence of this randomization, the various magnetization curves can be found here, especially for the sufficiently strong positive anisotropies $D^x$. In fact, in addition to the usually observed magnetization curves of the

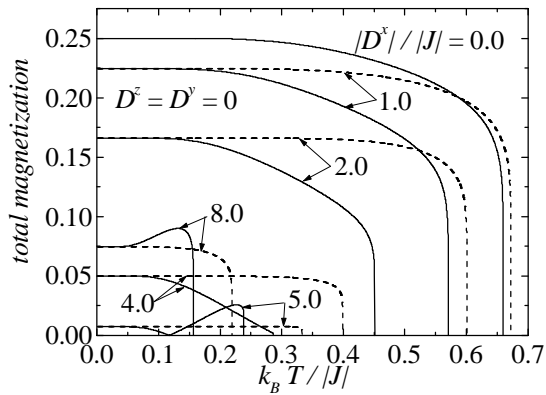

**Fig.5** Thermal variations of the total magnetization for different anisotropies $D^x$, when the anisotropy terms $D^z = D^y = 0$. The solid (dashed) lines corresponding to the relevant positive (negative) anisotropies $D^x$.

Q–type (for all negative and weak positive anisotropies $D^x$), the R–type magnetization curves ($D^x/|J| = 4.0$), N–type ($D^x/|J| = 5.0$) or P–type ($D^x/|J| = 8.0$) might be also realized. The occurrence of all the additional dependences is related to the fact, that the magnetization of the sublattice B is unsaturated and simultaneously thermally more easily disturbed than the magnetization of the sublattice A, which is saturated and thermally more stable (exhibits the Q–type behaviour irrespective of the crystal field strength $D^x$).

## 4. CONCLUSION

In the present article, we have investigated the mixed–spin Ising system in a presence of an anisotropic crystalline field by making use of an exact star–triangle mapping transformation. Our analysis has revealed that the behaviour of the considered system basically depends on the anisotropy terms $D^z - \Delta_1$ and $\Delta_2$. In fact, on the basis of the exact mapping one may conclude that the first term plays the role of an effective uniaxial anisotropy along the z–axis, whereas the second term (which occurs only when in–plane anisotropy $D^x \neq D^y$ is present) is responsible for the transverse–field like effect in the system. These results support the previous concept [7] that without loss of generality the most general anisotropy may be taken as: $D^x = E$, $D^y = -E$, $D^z = A$, where $E$ stands for the in–plane anisotropy and $A$ being the effective uniaxial anisotropy. Altogether, the most of the nontrivial results come from the analysis of the models with the nonzero in–plane anisotropy ($D^x \neq D^y$). In this case, the interesting quantum effects can be observed in the system. Namely, besides the standard phase transition behaviour (the system undergoes a second–order phase transition into the disordered phase), not perfectly spin ordering may be found in the system, even in the ground state. This behaviour may be interpreted as a competition between exchange interaction which tries to align the spins in the same direction and the effect of the anisotropy $\Delta_2$ which has tendency to destroy this alignment. Finally, it is worth noticing that our results are interesting from the theoretical point of view (because of the exactness of the applied method), as well as from the experimental point of view. Namely, a class of recently synthesized compounds $Ni(X)_2Ni(CN)_4$ [3] provides a clear experimental confirmation of the influence of the in–plane anisotropy, since the in–plane anisotropy significantly modifies the magnetic properties of this class of compounds. We hope, that the crystal field induced partial randomization theoretically predicted in this work, will be also experimentally confirmed in the near future. From this point of view, the most promising are the 2D systems in that strong Jahn–Teller effect causes the lattice distortion and consequently, an anisotropic ligand field (of lower symmetry) leads to the strong in–plane anisotropy in the crystalline field.

*Acknowledgment*: This work has been supported by the Ministry of Education of Slovak Republic under VEGA grant No. 1/9034/02.

## REFERENCES

[1] Siqueira AF and Fittipaldi IP,
    J. Magn. Magn. Mater. 54 (1986) 678.
[2] Gonçalves LL, Phys. Scripta 32 (1985) 248.
[3] Orendáč M et al, Phys. Rev. B 60 (1999) 4170;
    Orendáč M et al, Phys. Rev. B 61 (2000) 3223.
[4] Fisher ME, Phys. Rev. B 113 (1959) 969.
[5] Honmura R, Kaneyoshi T,
    J. Phys. C 12 (1979) 3979.
[6] Jaščur M, Lacková S,
    J. Phys.: Condens. Matter 12 (2000) L583.
[7] Papanicolaou N, Spathis PN,
    Phys. Rev. B 52 (1995) 16 001.

## BIOGRAPHY

Michal Jaščur was born on 16.10.1963. In 1987 he graduated (RNDr.) with distinction at the Department of Theoretical Physics and Geophysics of P. J. Šafárik University in Košice. He defended his PhD at alma mater in 1995. From 2000 he holds the position of the Associate Professor.
Jozef Strečka was born on 23.4.1977. In 2000 he graduated with distinction at the Department of Theoretical Physics and Geophysics of P. J. Šafárik University in Košice. At present, he is working as a PhD student at alma mater. The research interest of both authors are related to the quantum theory of magnetism and phase transition phenomena.